# Seismic Bayesian evidential learning: Estimation and uncertainty quantification of sub-resolution reservoir properties

Anshuman Pradhan[1] and Tapan Mukerji[2]
Department of Energy Resources Engineering
Stanford University, California, USA
Email: [1]pradhan1@stanford.edu, [2]mukerji@stanford.edu

## Abstract

We present a framework that enables estimation of low-dimensional sub-resolution reservoir properties directly from seismic data, without requiring the solution of a high dimensional seismic inverse problem. Our workflow is based on the Bayesian evidential learning approach and exploits learning the direct relation between seismic data and reservoir properties to efficiently estimate reservoir properties. The theoretical framework we develop allows incorporation of non-linear statistical models for seismic estimation problems. Uncertainty quantification is performed with Approximate Bayesian Computation. With the help of a synthetic example of estimation of reservoir net-to-gross and average fluid saturations in sub-resolution thin-sand reservoir, several nuances are foregrounded regarding the applicability of unsupervised and supervised learning methods for seismic estimation problems. Finally, we demonstrate the efficacy of our approach by estimating posterior uncertainty of reservoir net-to-gross in sub-resolution thin-sand reservoir from an offshore delta dataset using 3D pre-stack seismic data.

## 1 Introduction

Solving an inverse problem for elastic properties is a staple component of most seismic reservoir characterization (SRC) workflows. Elastic property inversion techniques typically encounter theoretical, numerical and computational complexities due to the high dimensional nature of subsurface models and the seismic data. It is a well-studied problem and there are many efficient approaches to tackle these challenges. However, in certain applications, especially related to reservoir management studies, the desired properties may be low-dimensional averaged properties over a reservoir interval like average net-to-gross or average fluid saturations. The research problem we pose and address in this paper is the following: Does the estimation of low dimensional reservoir properties from seismic data indispensably stipulate the solution of a complex high dimensional inverse problem?

The majority of the SRC workflows solve the seismic inverse problem in a causal framework by inferring each cause from its effect. Seismic data is the response of the elastic property model of the earth to the geophysical experiment and corresponding reservoir property earth model is linked to the elastic model through rock physics relations (Fig. 1). In contrast, when the final objective is low-dimensional properties instead of the full earth model, it would be preferable for the estimation strategy to entail the following: 1) Quantify the seismic signatures of the target properties in reduced dimensions 2) Solve the estimation problem in this reduced dimensional space. This alternative approach to estimation is the evidential learning approach. In this paper, we explore the efficacy of estimating reservoir properties from seismic data using this approach. A number of recent works, especially in reservoir performance forecasting [31, 39, 37, 38] have demonstrated the practical advantages of employing the evidential approach. Li [31] presents a general Bayesian framework for the evidential learning approach. The general recipe for Bayesian Evidential Learning (BEL) entails learning the statistical relationship between the target variables and the data (Fig. 1) with the help of a training set. Learning this relationship facilitates sampling the target posterior distribution within various Bayesian inference frameworks. We demonstrate that Approximate Bayesian Computation (ABC) [7, 9, 11, 43] allows efficient inference of average net-to-gross and average saturations from seismic data.

The introduction of notions of evidential learning to seismic estimation of average net-to-gross and fluid saturations can be attributed to Dejtrakulwong [15]. Interpretation of these properties in thinly interbedded sand-shale sequences was performed by relying solely on data-centric statistical kernel learning techniques, without resorting to explicit impedance inversion [16]. This paper extends on their work by casting the problem of reservoir property estimation in sub-seismic reservoirs into a rigorous BEL framework. Our investigations focus on the estimation of average net-to-gross (NtG) and average fluid saturations in sub-seismic reservoirs. We begin our analysis with a brief review of existing approaches to seismic reservoir property estimation. We formulate a rigorous theoretical framework for seismic BEL, which facilitates incorporation of unsupervised and supervised statistical/machine learning techniques for solving seismic estimation problems in a Bayesian framework. With the aid of synthetic examples and a real case application, we demonstrate various advantages of adopting the evidential approach for seismic estimation of low-dimensional properties.

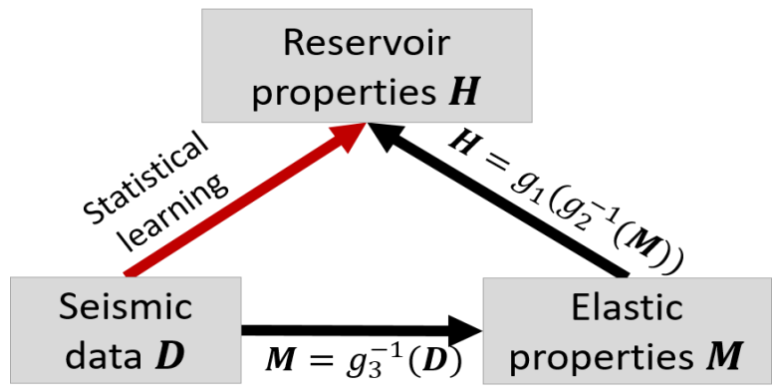

**Fig. 1** The causal analysis approach involves estimating each cause from its effect (black arrows). The evidential analysis approach performs estimation by learning the relationship between data and target quantities (red arrow) [31, 38]

## 2  Notation

We briefly describe the notation adopted for this paper. For any random variable $X$, $X \sim f_X(x)$ means $X$ is distributed according to the probability density function or distribution of $f_X(x)$. The corresponding lower-case letter $x$ denotes the sample values of the random variable $X$. Physical forward models are denoted by $g$, while regression models, parameterized by $\beta$, are represented as $q^\beta$. By low-dimensional reservoir properties, we generally refer to gross properties calculated across a subsurface depth interval such as net-to-gross (volume fraction of reservoir sand layers) or average fluid saturation. High-dimensional reservoir properties indicate the full geo-models of properties such as facies, porosity or fluid saturation. We represent the random vector (vector whose elements are random variables) for low and high dimensional reservoir properties by $\boldsymbol{H}$ and $\boldsymbol{H^h}$ respectively. Random vector for high-dimensional elastic properties, [P-wave velocity ($V_p$), S-wave velocity ($V_s$), bulk density ($\rho_b$)], are denoted by $\boldsymbol{M}$. Random vector for seismic data constituting pre-stack gathers are denoted by $\boldsymbol{D}$. Observed seismic data is denoted by $\boldsymbol{d_{obs}}$ and is viewed as a realization of $\boldsymbol{D}$. Letter $\boldsymbol{E}$ denotes the random vector for errors inherent in our estimation due to factors such as noise in data and forward modeling imperfections. $n_X$ denotes the dimensionality of $\boldsymbol{X}$.

## 3 Seismic Bayesian Evidential Learning

The theoretical framework for seismic BEL presented in this section is applicable for estimation of any reservoir property $\boldsymbol{H}$ implicitly related to seismic data given that $n_H \ll n_D$.

Given observed seismic data $\boldsymbol{d_{obs}}$, the goal of seismic reservoir property estimation in a Bayesian framework is to generate samples of the desired reservoir property from the posterior distribution $f_{\boldsymbol{H}|\boldsymbol{D}}(\boldsymbol{h}|\boldsymbol{D} = \boldsymbol{d_{obs}})$. $\boldsymbol{H}$ is generally a function of the full earth property model $\boldsymbol{H^h}$:

$$\boldsymbol{H} = g_1(\boldsymbol{H^h}) \tag{1}$$

For example, when $\boldsymbol{H}$ represents net-to-gross, $g_1(.)$ estimates the fraction of sand layers in the high dimensional facies vector $\boldsymbol{H^h}$. Traditional inversion methods would invert for $\boldsymbol{H^h}$ from $\boldsymbol{d_{obs}}$ and then extract $\boldsymbol{H}$ from $\boldsymbol{H^h}$. The BEL approach learns the relation between $\boldsymbol{H}$ and $\boldsymbol{D}$ from which it infers $\boldsymbol{H}$ conditioned to the observed data $\boldsymbol{d_{obs}}$.

The forward functional relation between $\boldsymbol{H^h}$ and $\boldsymbol{D}$ is generally specified as a synthesis of the rock physics model $g_2(.)$ and wave propagation model $g_3(.)$:

$$M = g_2(H^h) + E_2 \tag{2}$$

$$D = g_3(M) + E_3 \tag{3}$$

Here, $E_2$ and $E_3$ are error random vectors which account for effects not modeled such as noise in data and imperfections in the forward models $g_2(.)$ and $g_3(.)$ respectively. The conventional SRC workflows can be broadly categorized into sequential or simultaneous approaches based on how the above equations are inverted [13]. The sequential approach inverts Equations 2 and 3 in two cascaded steps, while the simultaneous approach solves a joint inverse problem. In the simultaneous approach, the desired posterior distribution is specified as below, using Bayes rule in terms of the prior distribution on $H^h$, rock physics likelihood $f_{M|H^h}(.)$ and seismic likelihood $f_{D|M}(.)$:

$$f_{H^h,M|D}(h^h, m|D = d_{obs}) \propto f_{D|M}(D = d_{obs}|m) f_{M|H^h}(m|h^h) f_{H^h}(h^h) \tag{4}$$

When $E_3$ is modeled as a Gaussian distribution $\mathcal{N}(0, \Sigma)$, the seismic likelihood is given as [45]

$$f_{D|M}(D = d_{obs}|m) \propto exp\left(-\frac{1}{2}(d_{obs} - g_2(m))^T \Sigma^{-1}(d_{obs} - g_2(m))\right). \tag{5}$$

The above approach provides an elegant and rigorous framework to invert for $M$, and $H^h$ from which $H$ can be obtained using Equation 1. However, if one considers a scenario where the goal is the estimation of just $H$ from $D$ and the direct relation between $D$ and $H$ is known, it would be possible to generate posterior samples of $H$, dispensing with the necessity for generating samples of either $H^h$ or $M$. The goal of BEL is to quantify this direct relationship between $H$ and $D$ by statistical learning and accomplish efficient lower-dimensional Bayesian inference.

The major steps, at the high level, in our workflow consist of:

1. Defining the prior distribution of the target variables and prior falsification;

2. Selection of an informative summary statistic and performing approximate Bayesian computation using the selected summary statistic; and finally,

3. Posterior falsification.

Below we describe these major components of BEL and propose methods suitable for seismic estimation problems.

## 3.1 Priors and prior falsification

The foremost step of BEL [40, 31], as with any Bayesian inference workflow, is to define the prior distributions on relevant variables. The priors, however, serve an additional purpose in BEL, that of generating the training set for the learning problem. Establishing a statistical model between $H$ and $D$ requires a training set consisting of the corresponding samples $\{(h_i, d_i); i = 1,..,l\}$, where $l$ is the number of samples in the training set. In most geoscientific experiments we typically have access to one realization $d_{obs}$ of $D$ and to a sparse subset (at wells) of the true earth realization $h^h_{true}$. In the BEL approach, we address this limitation by generating samples of $H^h$ from its prior distribution $f_{H^h}(h^h)$. The corresponding samples of $H$ and $D$ are generated by equations 1, 2 and 3. Following this approach, a large number of training examples, as suitable for the learning application, can be generated depending on the computational cost of the physical models. Note that the training examples are generated by sampling from the *prior* distribution which is a significantly easier problem than sampling techniques such as Markov Chain Monte Carlo (MCMC) or Gibbs sampling required for efficient exploration of high-dimensional *posterior conditional* probability spaces [45].

We emphasize that in generating the training samples of $D$, random vector $E$ for the forward modeling and data uncertainties should also be sampled and assimilated as specified in Equations 2 and 3. This in contrast to Bayesian inversion frameworks where the general approach is to assign the prior samples a likelihood value based on the probability distribution for the error (Equation 5), which is subsequently used in a sampling scheme to generate posterior samples. Note that in our case the statistical model will be employed for direct inference from real data. By sampling the error distribution during training set generation, we train the model to account for these uncertainties at prediction time.

We caution that the efficacy of our approach can be compromised by the issue of prior inconsistency, which is inherent in all Bayesian applications. Modeling of prior beliefs on uncertainty is highly subjective in nature and measures for validating the consistency of the prior with observed data are imperative. Mathematically, this is equivalent to establishing that $\boldsymbol{d}_{obs}$ and the training examples $\{\boldsymbol{d}_i; i = 1,..,l\}$ are realizations of the same random variable $\boldsymbol{D} \sim f_{\boldsymbol{D}}(\boldsymbol{d})$. The typical approach is to compare appropriate summary statistics of $\boldsymbol{d}_{obs}$ and training examples and determine if $\boldsymbol{d}_{obs}$ is an outlier. The outlier detection algorithm employed depends on the characteristics and dimensionality of the $\boldsymbol{d}_{obs}$. Li [31] discusses some methods to perform such tests and associated challenges. In the real case application presented in a later section, outlier detection using Mahalanobis distances is employed to determine if the statistics of the observed seismic angle gather is captured by the prior samples of the training set. In the eventuality that $\boldsymbol{d}_{obs}$ is deemed to be an outlier, then the prior distributions will be falsified, necessitating suitable adjustments to the prior, either by broadening the range of parameters, or by substantially changing the geological scenario.

### 3.2 Strategies for seismic BEL

The efficacy of BEL is critically dependent on selecting a statistical learning method that facilitates effective and efficient Bayesian inference. To this end, previous applications of BEL [31, 37] have advocated learning by linear and non-parametric regression techniques. Regressing $\boldsymbol{D}$ on $\boldsymbol{H}$ with the linear model $q^\beta$,

$$\widehat{\boldsymbol{D}} = q^\beta(\boldsymbol{H}) \tag{6}$$

facilitates analytical formulation of $f_{\boldsymbol{H}|\boldsymbol{D}}(\boldsymbol{h}|\boldsymbol{D} = \boldsymbol{d}_{obs})$ if $f_{\boldsymbol{H}}(\boldsymbol{h})$ is assumed to be Gaussian. Here, $q^\beta$ represents a regression model parameterized by $\beta$ and $\widehat{\boldsymbol{D}}$ is the estimated data variable. Non-parametric regression methods [41, 42] relax the strong assumptions of linearity and Gaussianity by fitting kernel functions to estimate $f_{\boldsymbol{H}|\boldsymbol{D}}(\boldsymbol{h}|\boldsymbol{D} = \boldsymbol{d}_{obs})$, but are normally applicable in low-dimensional settings. In our case the seismic data $\boldsymbol{D}$ is high dimensional. Even though these methods facilitate efficient sampling of the posterior, they are not conducive to seismic problems. The direct relation between $\boldsymbol{D}$ and $\boldsymbol{H}$ cannot

be expected to be linear in general. Additionally, while $H$ may be low dimensional, the high dimensional nature of seismic data precludes non-parametric regression methods. An obvious solution would be to employ regression of $D$ on $H$ with non-linear $q^\beta$ and subsequently sample the posterior by standard MCMC techniques. However, we do not recommend this approach as obtaining a reliable regression model will be challenging since $n_H \ll n_D$.

To address the various issues highlighted above, we propose the following approach for applying BEL in seismic applications:

1. Extract informative features from $D$ which quantify the non-linear relationship with $H$ as desired. These features are referred to as summary statistics of the data and denoted as $S(D)$.
2. Account for $S(D)$ during Bayesian inference, i.e., sample the distribution $f_{H|S(D)}(h|S(D) = S(d_{obs}))$.

However, since $S(.)$ could be a non-linear function, it presents complications for formulating an exact model for the likelihood. Recall from Equation 5 that the likelihood distribution is assigned according to the data and modeling uncertainty, specified typically as $E \sim \mathcal{N}(0, \Sigma)$. It is difficult to derive a probabilistic model for the noise following application of an arbitrary non-linear transformation $S(.)$ of the data variable. While it is possible to fit probabilistic models to $S(E)$ using Monte Carlo sampling [24], such approach requires making assumptions about the parametric nature of the model. Rather, we advocate likelihood-free inference using Approximate Bayesian Computation (ABC) as described in the sub-section below. Details on selecting an informative summary statistic $S(D)$ using unsupervised and supervised learning is presented in the subsequent sub-section.

## 3.3 Inference by Approximate Bayesian Computation

ABC was developed as a framework for performing Bayesian inference in problems which exhibit intractable likelihoods [7, 9, 43]. Equation 7 captures the essence of ABC by showing the series of approximations necessary in order to sample the posterior in the absence of a tractable likelihood.

$$f_{H|D}(h|D = d_{obs}) \approx f_{H|S(D)}(h|S(D) = S(d_{obs}))$$

$$\approx f_{H|S(D)}(h| \parallel S(d) - S(d_{obs}) \parallel < \epsilon_S) \qquad (7)$$

Here, $\parallel \cdot \parallel$ is a distance measure and $\epsilon_S$ is a threshold dependent on $S(.)$. Theoretically, it is possible to sample the exact posterior (L.H.S. of Equation 7) using rejection-sampling by generating prior samples $\{h_i; i = 1,..,n\}$, obtaining corresponding data $\{d_i; i = 1,..,n\}$ and accepting those $h_i$s for which $d_i = d_{obs}$. Since the probability of finding an exact match will generally be very low, one instead hopes to find a match to some low-dimensional summary statistics $S(d_{obs})$ (middle expression of Equation 7). In practice, the distribution specified on the R.H.S. of Equation 7 is sampled by accepting models for which $\parallel S(d_i) - S(d_{obs}) \parallel < \epsilon_S$. The threshold $\epsilon_S$ is set such that the approximation holds true. In the limiting case where $S(.)$ is the identity function and $\epsilon_S = 0$, equation 7 becomes trivial. We highlight that ABC, by allowing to incorporate any desired summary statistics into the analysis, serves as the natural inference framework for seismic BEL applications.

### 3.4 Selecting an informative summary statistic

A small threshold $\epsilon_S$ improves the approximation in Equation 7 but requires a large number of forward models to generate posterior samples that meet the threshold. Employing an informative summary statistic can help maintain the accuracy and efficiency of the approach [10, 11], by allowing Equation 7 to hold even for a relatively larger value of $\epsilon_S$. The ideal summary statistic should: 1) provide a highly compressed representation of $D$ with minimal loss of information 2) be highly informative on the target variable $H$. We propose the following approach to select a suitable statistic: 1) Depending on the nature and dimensionality of $D$ and $H$, formulate a set of potentially informative summary statistics; 2) Generate a test example $(h, d)$ by sampling from the prior distribution and forward modeling the data; 3) Compare the posterior distributions estimated by performing ABC using each statistic. Summary statistics highly informative on $H$ will exhibit significant reduction of the prior uncertainty in the corresponding posterior distributions. For summary statistics that are not informative about $H$, the posterior and prior distributions of the target variable will be similar.

To formulate a set of potential $S(.)$, we consider statistics learned in unsupervised as well as supervised settings. Unsupervised learning methods reduce data dimensionality by eliminating redundancy in the data, extracting features capturing patterns present in the data and learning lower-dimensional manifolds along which data might be distributed [8]. These methods could be effective for our problem since $n_H \ll n_D$ and thus the seismic data space might contain implicit low-dimensional representations capturing the variability of $H$. We also consider supervised learning methods that extract features by explicitly encoding the relationship between $D$ and $H$ using the training set. Specifically, $H$ is regressed on $D$ and the resulting estimator is used as a summary statistic:

$$S(D) = \widehat{H} = q^\beta(D) \qquad (8)$$

where the regression model $q^\beta$ is learned by minimizing the loss function shown below, given the training set $\{(h_i, d_i);\ i = 1,..,l\}$,

$$\mathcal{L}(H, \widehat{H}) = \frac{1}{l}\sum_{i=1}^{l} \|\widehat{h}_i - h_i\|_2^2 = \frac{1}{l}\sum_{i=1}^{l} \|q^\beta(d_i) - h_i\|_2^2 \qquad (9)$$

Supervised learning methods require an optimization problem to be solved, are more computationally intensive and more prone to overfitting than many unsupervised methods. However, summary statistics extracted through supervised learning have several advantages: 1) Since $S(D) = \widehat{H}(d)$, it directly informs on $H$ 2) The degrees of freedom of the statistic is low as $H$ is low dimensional and thus allows us to set a relatively high value for $\epsilon_S$. 3) Using $\widehat{H}$ as a summary statistic in an ABC framework has the attractive theoretical result that the approximate and exact posterior converge in expectation [21]. Explicit examples of both unsupervised and supervised methods to obtain summary statistics are described in the section 4.

## 3.5 Falsification of the approximate posterior

Equation 7, albeit facilitating generation of desired samples, renders our approach susceptible to mis-approximations. These could stem from various factors such as: 1) matching data summary statistics might be a poor approximation to matching the data itself 2) Threshold $\epsilon_S$ might not be small enough. To ensure reliability, it is imperative to determine whether a particular summary statistic and associated threshold reasonably approximate Equation 7. We present below a method of falsifying the approximate posterior to establish this result. For brevity, we denote the approximate posterior $f_{H|S(D)}(h| \parallel S(d) - S(d_{obs}) \parallel < \epsilon_S)$ as $\widetilde{f_{H|D}}(h|d_{obs})$ in the following.

Consider the property $h_{true}$ of the true earth model and observed data $d_{obs}$. We treat the true earth property and observed data as random vectors $H_{true}$ and $D_{obs}$ respectively. The approximate posterior will be falsified if equation 10 and consequently equation 11 do not hold true.

$$H_{true} \sim \widetilde{f_{H|D}}(h|d_{obs}) \tag{10}$$

$$\Rightarrow p\left(H^j_{true} \leq P_\delta\left(\widetilde{f_{H^j|D}}(h^j|d_{obs})\right)\right) = \delta\% \; ; \forall j, \delta \tag{11}$$

Here, $p(.)$ denotes probability, superscript $j$ denotes the $j^{th}$ element of a vector and $P_\delta(.)$ is the $\delta\%$ quantile of a distribution.

In other words, if $\widetilde{f_{H^j|D}}(h^j|d_{obs})$ reasonably approximates the distribution of $H^j_{true}$, then the probability that $H^j_{true}$ is less than any quantile $P_\delta$ of $\widetilde{f_{H^j|D}}(h^j|d_{obs})$ should be $\delta\%$. The posterior falsification step tests whether equation (11) holds. To test this, we generate a new test set $\{(h_{true,i}, d_{obs,i}); i = 1, .., n\}$, consisting of realizations of $H_{true}$ randomly sampled from the prior and corresponding forward-modeled samples of $D_{obs}$. Given $i^{th}$ observed data sample $d_{obs,i}$, we estimate corresponding approximate posterior distribution using equation 7 and estimate $P_\delta(\widetilde{f_{H^j|D}}(h^j|d_{obs,i}))$. This analysis is conducted for all test set examples. Subsequently, L.H.S. of equation 11 is empirically

evaluated by counting the fraction of test samples in which $h^j_{true,i}$ is less than the corresponding $P_\delta(.)$ as shown below:

$$p\left(H^j_{true} \leq P_\delta\left(\widetilde{f_{H^J|D}}(h^j|\boldsymbol{d}_{obs})\right)\right) = \frac{\sum_{i=1}^{n}\mathbb{I}(h^j_{true,i} \leq P_\delta(\widetilde{f_{H^J|D}}(h^j|\boldsymbol{d}_{obs,i})))}{n} \quad (12)$$

Here, $\mathbb{I}(.)$ denotes the indicator function. Note that this result should hold true for all variables $H^j$ and quantiles $P_\delta(.)$ to validate equation 10.

## 4 A synthetic example

In this example, we assume that the reservoir interval is 200 meters thick, at a depth of 1500 meters with a shaly overburden. The reservoir interval contains thin sand layers below seismic resolution, interbedded with shaly sand and shale. The goal is to use the seismic BEL approach for estimating the reservoir net-to-gross and average fluid saturations in the reservoir interval using near and far offset seismic data. We will evaluate the efficacy of our approach using a test set consisting of example pairs of true synthetic earth models and observed seismic gathers, generated by sampling of the prior distributions and seismic forward modeling as discussed below. For the purposes of this synthetic example, we assume that the prior distributions and the forward modeling scheme generating the observed data are known to us. Since the priors are consistent with the observed data, falsification of the prior is not required in this case. The real field data case described later will require this step.

### 4.1 Priors for facies, elastic properties, and seismic data

We assume that the prior uncertainty on the vertical spatial arrangement of the reservoir facies layers can be captured by a Markov chain model. Transition matrices for Markov chains describing lithologic successions are typically estimated using well logs [20]. The three-state transition matrix we use [23] for our example is shown in Table 1. The three states correspond to sand, shaly sand and shale lithologies. Realizations are generated by sampling the Markov chain with a constant vertical discretization of 1 meter. Realizations of the prior facies model are shown in Fig. 2. We assume that priors on the elastic properties are described by facies-conditional Gaussian distributions: $f_{M^i|H^{h,i}}(m^i|h^{h,i}) \sim \mathcal{N}(\mu, \Sigma)$. Here, $M^i$ ($\in \mathbb{R}^3$) denotes the random variable for elastic properties, ($V_p$, $V_s$ and $\rho_b$) and $H^{h,i}$ is the random variable for facies respectively at depth $z^i$, $\{i = 1501, 1502, \ldots, 1700 \text{ meters}\}$.

**Table 1** The transition matrix for the Markov chain of the synthetic example

|  | Sand | Shaly sand | Shale |
|---|---|---|---|
| Sand | 0.9 | 0.05 | 0.05 |
| Shaly sand | 0 | 0.93 | 0.07 |
| Shale | 0.05 | 0 | 0.95 |

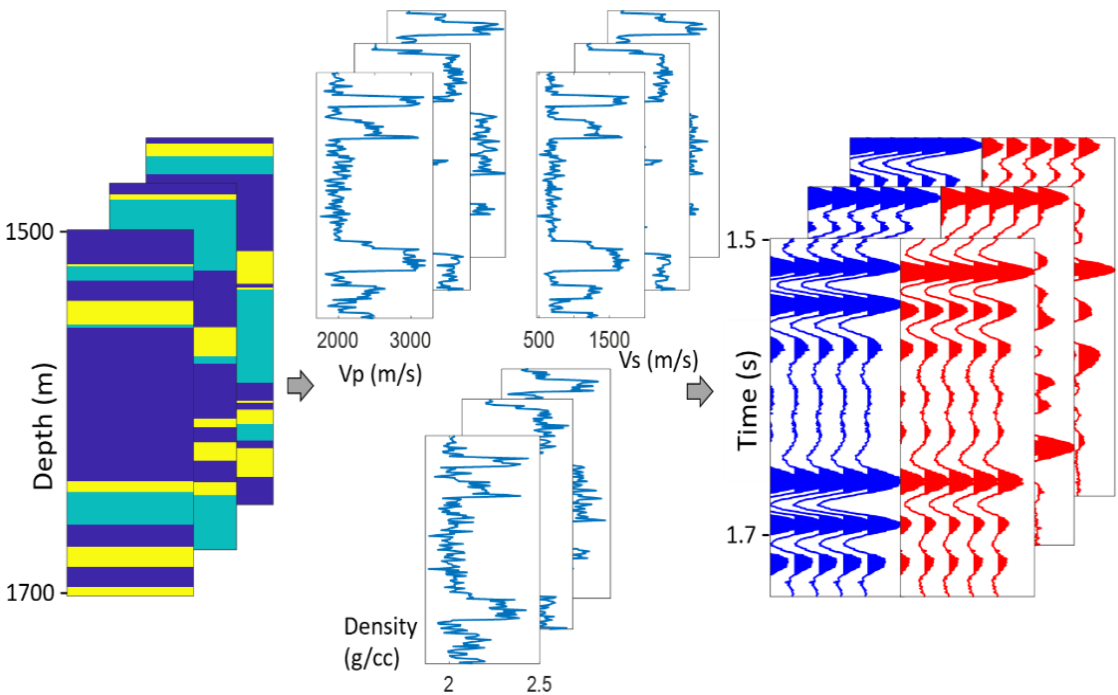

**Fig. 2** (From left to right) Prior realizations of facies (sand: yellow, shaly sand: green and shale: blue), elastic properties and forward modeled normal incidence (blue) and far angle trace (red). Each angle trace is repeated 5 times for visual convenience.

We use standard rock physics models to obtain the relation between the three facies and their corresponding mean $V_p$, $V_s$ and $\rho_b$. We use a combination of the Yin-Marion dispersed-mixing model [33] and the soft-sand rock physics model [5] as described by Dejtrakulwong [15, Chapter 4] to obtain $\boldsymbol{\mu}$. For all facies, $\boldsymbol{\Sigma}$ is assumed to be:

$$\boldsymbol{\Sigma} = \begin{bmatrix} \sigma_{V_p}^2 & r\sigma_{V_p}\sigma_{V_s} & r\sigma_{V_p}\sigma_{\rho_b} \\ r\sigma_{V_p}\sigma_{V_s} & \sigma_{V_s}^2 & r\sigma_{V_s}\sigma_{\rho_b} \\ r\sigma_{V_p}\sigma_{\rho_b} & r\sigma_{V_s}\sigma_{\rho_b} & \sigma_{\rho_b}^2 \end{bmatrix} \quad (13)$$

Here, $\sigma_{V_p} = 100$ m/s, $\sigma_{V_s} = 70$ m/s, $\sigma_{\rho_b} = 0.05$ g/cc and $r = 0.8$. Kernel density estimates of the facies-conditional priors distributions for $V_p$, $V_s$ and $\rho_b$ as well as the bivariate facies-conditional distributions of P-impedance ($I_p = V_P\rho_b$) and S-impedance ($I_s = V_S\rho_b$) are shown in Fig. 3. Two different fluid-saturation scenarios are considered as described below.

1. *Scenario 1*: All facies are assumed to be completely water saturated and the goal in this scenario is to estimate only NtG over the reservoir interval.

2. *Scenario 2:* We assume that the pore fluid in the sand layers consists of oil and water mixed in unknown proportions, while other facies are completely water saturated. The goal in this case is to estimate both NtG and average water saturation in the reservoir interval. We assume a uniform prior $\mathcal{U}(0.15,1)$ on the water saturation value in each sand layer. The effect of fluid substitution on the elastic properties is derived using Gassmann's [34] fluid substitution relations.

Single-scattering forward modeling with the exact non-linear Zoeppritz equation [4], 35 Hz Ricker wavelet and 1% independent and identically distributed Gaussian noise (signal to noise ratio SNR = 100) was employed for generation of the normal incidence and far angle ($30^0$) seismic traces as shown in Fig. 2. $\boldsymbol{D}$ consists of both these traces and is 840 dimensional with a time discretization of 0.5 ms. Note that the sand layers in the reservoir facies realizations will be sub-seismic with wavelength to average layer thickness ratio $\lambda/d_{avg} \approx 7$. The Markov Chain has a depth discretization $dz = 1$ meter and sand layers have a self-transition probability $p = 0.9$, thus $d_{avg} = \frac{dz}{(1-p)} = 10$ meters [20]. $\lambda$ has a value of $\approx 70$ meters corresponding to mean sand $V_p$ of 2425 m/s.

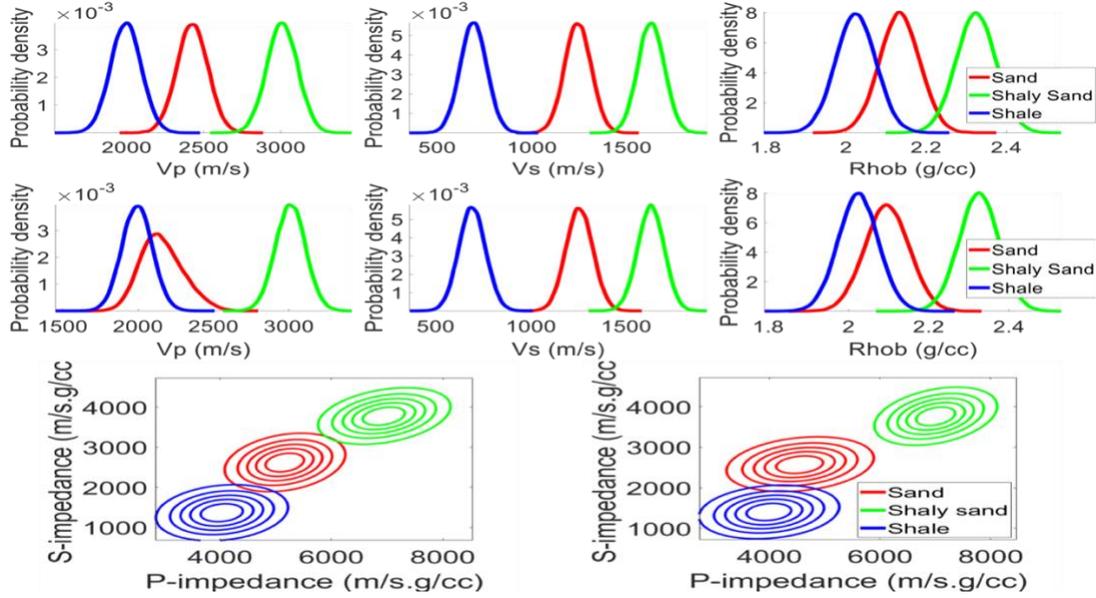

**Fig. 3** Kernel density estimates of prior facies-conditional distributions of $V_p$, $V_s$ and $\rho_b$ in synthetic case scenarios 1 (top row) and 2 (milddle row). In the bottom row, bivariate distributions in the $I_p - I_s$ space are shown for scenarios 1 (left) and 2 (right)

### 4.2 Summary statistic synthesis: Unsupervised learning methods

We explored 3 different dimension reduction methods to identify informative summary statistics extracted in an unsupervised manner. Training and validation sets (examples 'hidden' from the training process) consist of 50000 and 2000 examples of seismic data ($\boldsymbol{D} \in \mathbb{R}^{840}$) respectively. Results are shown for Scenario 1, estimating NtG only, as it turned out that the unsupervised methods explored did not give sufficiently informative statistics.

1. *Principal Components Analysis (PCA)*: PCA is one of the simplest and most effective dimension reduction techniques available. PCA helps eliminate redundancies present in the data by identifying orthogonal directions (principal components, in the data space, which maximize the variance of data. In the event that majority of the data variance is captured along a few principal components, dimension reduction can be achieved by discarding the other components. Using the training set, 49 out of 840 principal components were retained which capture approximately 99% of the training data variance. In Fig. 4, we show the original seismic trace for a validation set example and the trace reconstructed from the 49-dimensional principal component space identified using the training set. The metrics reported in Table 2 indicate excellent agreement between original and reconstructed seismic traces for all validation set examples.

2. *Wavelet thresholding/shrinkage*: A limitation of PCA is that the principal components are obtained through linear transformations of the original data dimensions and might not capture non-linear features present in the data. To address this limitation, we used wavelet thresholding, which has been shown to be quite effective in denoising, compressing and finding structures in data [18, 19, 46]. The general approach is to perform a discrete wavelet transform (DWT) [32] of the data and apply a thresholding rule to retain only significant wavelet coefficients, thus obtaining a sparse representation of the data. We employ the training set to determine the wavelet functions which maximally capture the variations in our data. A 4-level DWT using Daubechies least asymmetric wavelet bases [14] is performed on all the training examples to obtain corresponding coefficients of the scaling and wavelet functions. We retain all the scaling functions and only 5% of the wavelet functions having highest coefficients after averaging across the training set. This approach reduces data dimensionality from 840 to 99. To evaluate the robustness of the chosen wavelet functions, we reconstruct the seismic traces in the training and validation set by performing inverse DWT on respective thresholded wavelet coefficients. Fig. 4 compares the original and reconstructed traces for a validation set example. Table 2 summarizes the results for all training and validation set examples.

3. *Compression using autoencoders*: Autoencoders (non-linear PCA) provide a framework for performing parametric non-linear feature extraction from the data [25, 26]. The parametric form of the features is embodied in an encoder-decoder neural network architecture. The encoder extracts low-dimensional features from the input through multiple non-linear hidden layers, while the decoder reconstructs the input from these features. The network weights are learned by minimizing the reconstruction error on the training examples. Note that learning features with autoencoder will reduce to performing PCA if all the hidden layers are linear. A factor which renders autoencoders potentially powerful is that they allow regularizing the training via various techniques which can prove effective in learning the underlying data-generating distribution [8]. We trained an autoencoder with a single hidden layer of 100 nodes and non-linearity was imposed through the sigmoid function:

$$f(x) = \frac{e^x}{e^x+1} \tag{14}$$

The training was regularized using $\ell_2$ and sparsity regularization. Sparsity in the learned features is encouraged by penalizing deviation of average activations of the hidden layer, averaged across the training set, from a low value. This causes each hidden layer node to get activated by distinctive features present in a small number of training examples [30]. Regularization coefficients, which define the weight assigned to corresponding regularization scheme in the loss function, were tuned to obtain good performance on the validation set. Performance of the trained autoencoder on the validation set (see Table 2 and Fig. 4) evidences good generalization of the learned features to unseen examples.

4. *Seismic waveforms*: For comparison, we also consider directly using the uncompressed seismic amplitudes as the summary statistic: i.e., $S(\boldsymbol{D}) = \boldsymbol{D}$.

**Table 2** Values for and % variance captured in the reconstructed seismic traces using unsupervised methods as against the original and correlation coefficients (CCs). Values are averaged across each evaluation set

| Evaluation set | PCA | Wavelet thresholding | Autoencoders |
|---|---|---|---|
| **Training set** | Variance captured: 98.99%, CC: 0.99 | Variance captured: 98.95%, CC: 0.99 | Variance captured: 97.89%, CC: 0.99 |
| **Validation set** | Variance captured: 98.98%, CC: 0.99 | Variance captured: 98.94%, CC: 0.99 | Variance captured: 97.86%, CC: 0.99 |

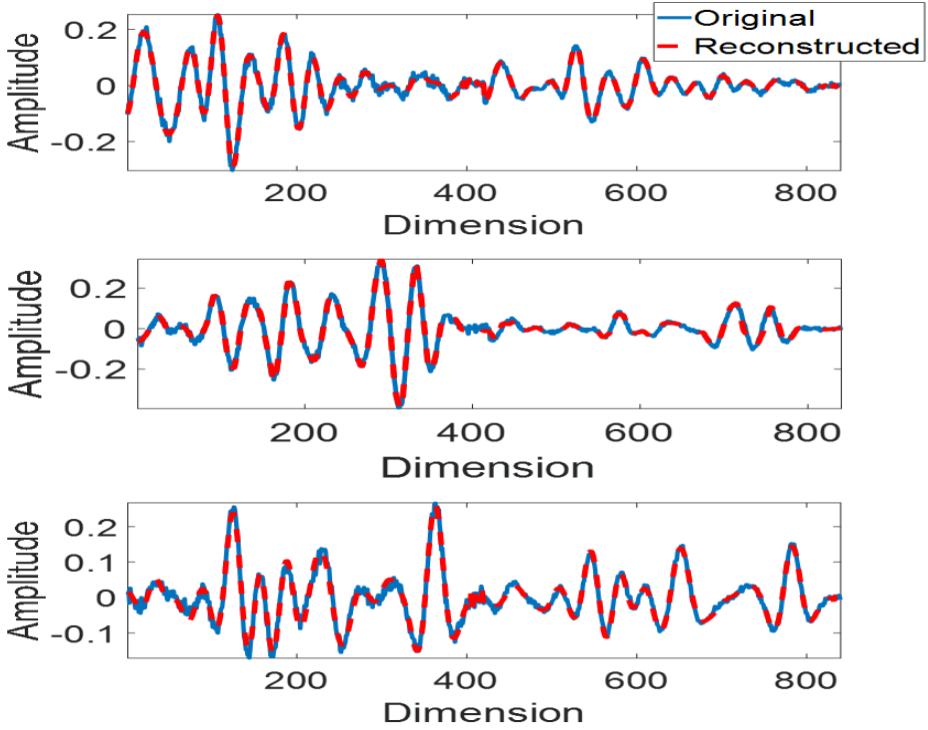

**Fig. 4** Original and reconstructed seismic traces obtained using PCA (top), wavelet thresholding (middle) and autoencoders (bottom) for different validation set examples

Equation 7 is utilized to estimate the approximate posterior for a randomly chosen $\boldsymbol{d}_{obs}$ from the test set (Fig. 5). We recommend excluding the training examples from the prior sample set on which ABC is executed since factors such as overfitting of the statistical learning method to the training examples might bias the resulting uncertainty estimates. We sample a separate batch of $1.5 \times 10^5$ prior models for the analysis. Given a summary statistic $\{S^k(.); k = 1,..4\}$ from the four unsupervised methods listed above, the following steps are carried out:

1. Calculate the summary statistics for all $l = 1.5 \times 10^5$ prior samples: $\{S^k(\boldsymbol{d}_1), S^k(\boldsymbol{d}_2), \dots, S^k(\boldsymbol{d}_l)\}$

2. Calculate the $\ell_2$-distance between the summary statistics of prior samples and observed data: $\{\| S^k(\boldsymbol{d}_i) - S^k(\boldsymbol{d}_{obs}) \|_2, i = 1,.., l\}$

3. From the prior samples $\boldsymbol{h}_i$s, accept those as posterior samples for which $\| S^k(\boldsymbol{d}_i) - S^k(\boldsymbol{d}_{obs}) \|_2 < \epsilon_S$. The threshold $\epsilon_S$ is specified using the approach

suggested by Beaumont et. al. [6], in which a certain percentage $\delta$ of the prior samples with the lowest $\| S^k(d_i) - S^k(d_{obs}) \|_2$ are accepted. Note that this corresponds to setting $\epsilon_S$ as the $\delta\%$ quantile of the empirical distribution for $\| S^k(D) - S^k(d_{obs}) \|$. The threshold value is set using posterior falsification, described in a later section.

4. Estimate the approximate posterior $f_{H|S(D)}(h| \| S^k(D) - S^k(d_{obs}) \|< \epsilon_S)$ empirically using the kernel density estimate (KDE) of the samples accepted in the previous step.

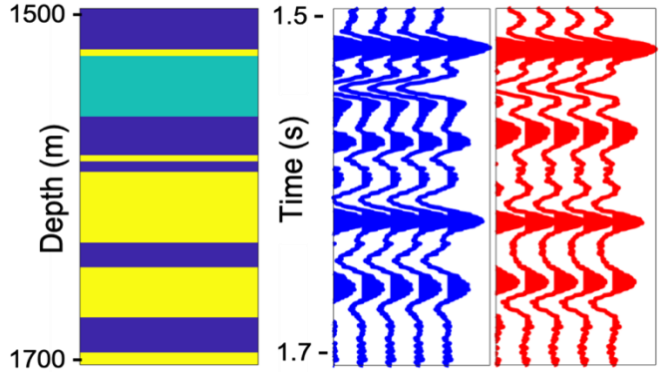

**Fig. 5** Randomly selected example from the test set on which synthetic case is tested. True facies model (left), normal incidence (blue) and far angle (red) traces are shown

A very low ($\delta = 0.1\%$) threshold for ABC was set to accept 150 prior models as the posterior samples. Comparison of the prior and posterior KDEs (Fig. 6) shows negligible reduction of prior uncertainty in all the cases. Possible factors responsible for this behavior could be that the summary statistics are not informative on the property of interest or the threshold is too high. It is easy to analyze the validity of the threshold when $S(D) = D$, since the likelihood can be expressed in an analytical form in this case. The posterior samples should fit $d_{obs}$ within the uncertainty of the 1% independent and identically distributed Gaussian noise ($E_3 \sim \mathcal{N}(0, \Sigma_d)$) added to the seismic gathers in test set. However, we found that none of the $1.5 \times 10^5$ prior models satisfy this criterion. Fig. 7 elaborates this observation by illustrating the KDE of the negative log likelihood, $\frac{1}{2}(d_{obs} - g_3(m_i))^T \Sigma_d^{-1}(d_{obs} - g_3(m_i))$, computed using all prior models $\{m_i; i = 1,..,l\}$. We also generated $l$ samples of $E_3$ and show the KDE of the corresponding negative log distribution values (i.e. $\frac{1}{2}\epsilon_3^T \Sigma_d^{-1} \epsilon_3$). The fact that the support of the two KDEs is significantly different proves threshold of $\delta = 0.1\%$ is too high when $S(D) = D$. Generating valid posterior samples will necessitate further random prior sampling or designing smart strategies for exploring the prior model space, like MCMC and Gibbs sampling, as is the norm in high-dimensional property inversion. For the purposes of low-dimensional property estimation, however, we emphasize that the efficiency of the ABC rejection sampling approach can be maintained if the employed summary statistic is highly informative on target properties. The previous analysis demonstrated that the evaluated unsupervised learning techniques are not sufficiently informative on NtG. Rather than using purely unsupervised, data-centric summary statistics, we propose to encode information about dependencies between data and the target variables into the summary statistic using supervised learning models.

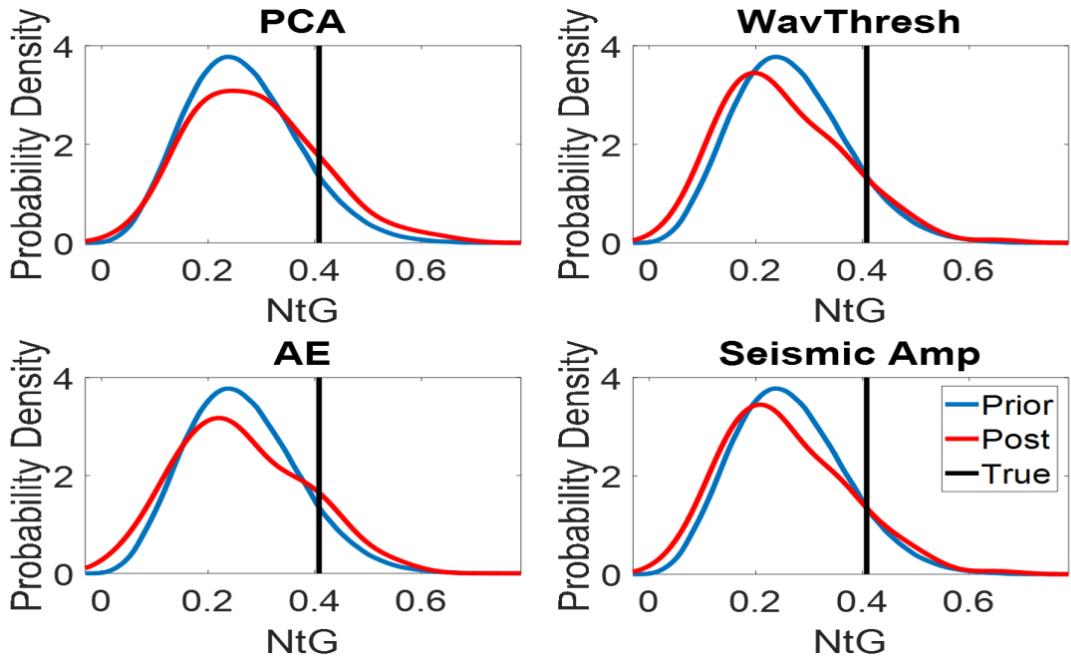

**Fig. 6** Estimated posterior distributions of NtG (red) estimated using different summary statistics as listed on top of each plot. The prior distribution (blue) and the true values (black) are shown for comparison

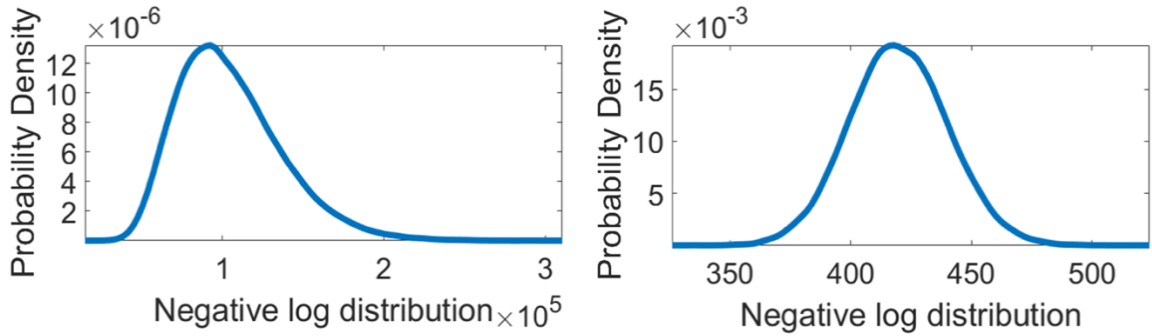

**Fig. 7** (Left) KDE of the negative log likelihood values obtained using $1.5 \times 10^5$ prior models (Right) KDE of negative log distribution using $1.5 \times 10^5$ samples from the error model

### 4.3 Summary statistic synthesis: Supervised learning using deep neural networks

The usage of various regression models of the target variable as summary statistic for ABC has been studied by several authors [21, 28, 47]. Similar to unsupervised methods, numerous supervised regression model architectures exist in literature today. While features learned by unsupervised methods might vary from one method to another, the characteristics of features learned by supervised methods, i.e. $\widehat{H}$, will remain same irrespective of the statistical model employed. The singular constraint is that the model should have the architectural flexibility to allow it to robustly learn the implicit relationship between data and target variables, thus exhibiting reliable prediction ability on examples extraneous to the training set. Deep neural networks (DNNs) is the regression model of choice for our problem. Our choice was motivated by the fact that the recent advances in deep learning research have made it feasible to train very deep networks with large number of learnable parameters without overfitting. Specific regularization techniques, such as dropout, have been shown to be highly effective in tackling overfitting.

Fig. 8 depicts our DNN architecture for Scenario 1. The network takes as input the vector consisting of the seismic gather and outputs NtG. The relationship between $D$ and $H$ is modeled using three fully connected hidden layers containing two leaky Rectified Linear

Unit non-linearities (equation 15) and a sigmoid activation function respectively (equation 14).

$$f(x) = \begin{cases} x & if\ x > 0 \\ 0.01x & otherwise \end{cases} \quad (15)$$

Note that the learnable parameters $\beta$ of the network consist of the filter weights of the hidden layers. For the second scenario, when estimating both NtG and saturation, only the output dimension of the network is modified from one output to two. The network was trained with Adam optimizer [29] in *tensorflow* deep learning framework.

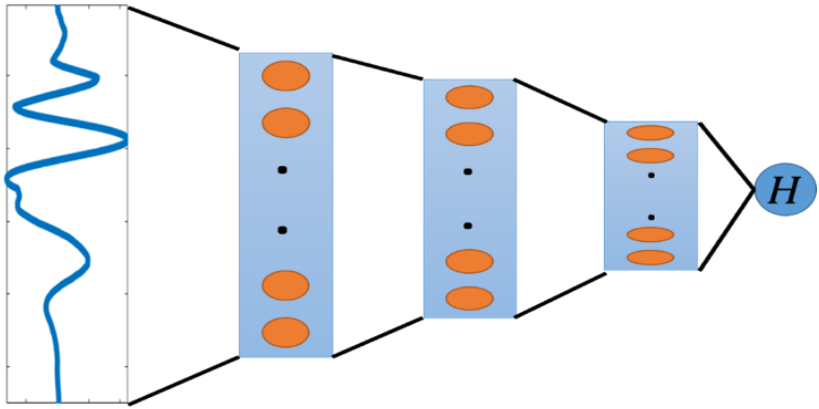

**Fig. 8** The deep neural network architecture

Several diagnostic measures were employed to ensure that the model is not overfitting and is able to generalize to previously unseen examples: 1) Performance on validation sets is evaluated simultaneously during training to identify overfitting, if any. 2) Regularizing the network weights during training aids in better generalization of the network's prediction ability. Regularization combining dropout [44] and batch-normalization [27] was found to be particularly effective for our network. Various hyper-parameters related to the network architecture, optimization algorithm and regularization scheme also require to be assigned. Suitable values (Table 3) were chosen by optimizing the network's performance on the validation set.

Table 3 Optimized hyper-parameters for DNN

| Case | Hidden layer sizes | Regularization scheme | Dropout rate | Mini batch size | Learning rate |
|---|---|---|---|---|---|
| **Synthetic scenario 1** | [708,446,143] | Dropout + Batch Normalization | 0.19 | 256 | 0.0007 |
| **Synthetic scenario 2** | [685,378,146] | Dropout + Batch Normalization | 0.32 | 512 | 0.0009 |
| **Real Case** | [629,387,77] | Dropout + Batch Normalization | 0.32 | 256 | 0.0008 |

Table 4 DNN training details

| Case | Training/Validation set sizes | Training epochs | Run time (on 1 NVIDIA K80 GPU) |
|---|---|---|---|
| **Synthetic scenario 1** | 50000/2000 | 800 | $\approx$ 30 minutes |
| **Synthetic scenario 2** | 90000/5000 | 800 | $\approx$ 60 minutes |
| **Real Case** | 100000/5000 | 100 | $\approx$ 10 minutes |

Table 5 Performance metrics of trained DNN. CC and RMSE refer to correlation coefficient and root mean square error between $\hat{H}$ and $H$

| Evaluation set | Synthetic scenario 1 | Synthetic scenario 2 | | Real Case |
|---|---|---|---|---|
| | NtG | NtG | Sw | NtG |
| **Training set** | CC: 0.99, RMSE: 0.02 | CC: 0.97, RMSE: 0.03 | CC: 0.96, RMSE: 0.01 | CC: 0.86, RMSE: 0.04 |
| **Validation set** | CC: 0.89, RMSE: 0.05 | CC: 0.87, RMSE: 0.05 | CC: 0.87, RMSE: 0.02 | CC: 0.81, RMSE: 0.05 |

The training details and performance metrics for the two estimation scenarios are listed in Table 4 and Table 5 respectively. In both scenarios, training was executed for 800 epochs and the network weights for the epoch with best validation set performance were retained as the final model. Fig. 9 depicts performance of Scenario 1's network on training and validation sets. The good performance on the validation set of 2000 examples signifies the reliability of the trained network. A larger training set was required to ensure reliable performance on the validation set for Scenario 2 (90000 vs. 50000 for Scenario 1). This observation can be attributed to the fact that fluid effects create more overlap between sand and shale $V_p$ values (Fig. 3), which potentially limits the network's ability to generalize easily. Note however that the prior distributions for the facies are well demarcated with respect to $V_s$ values. The afore-mentioned training behavior of the DNN indicates that it was able to learn this underlying distribution with additional training examples. Thus, in problems where forward modeling computational expense is not a limitation, BEL's approach to synthetic training set generation can prove effective in boosting the network's generalization ability by increasing the number of training examples as required.

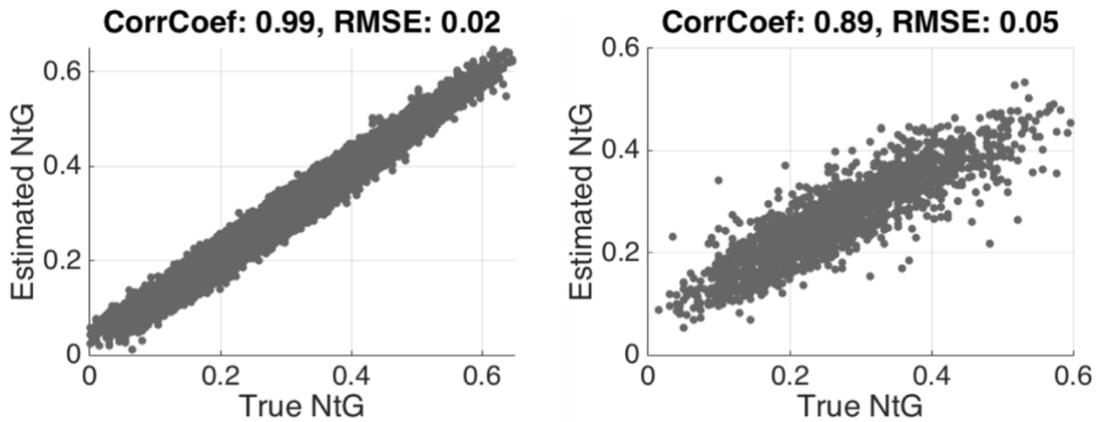

**Fig. 9** Plot of actual NtG of Scenario 1 against the NtG estimated by the trained network for the training (left) and validation (right) set. The correlation coefficient and RMSE between true and estimated values are reported at the top

To generate samples of the approximate posterior, we use DNN output $\widehat{H}$ as the summary statistic in the ABC workflow presented earlier. For performing ABC, $10^4$ prior models were sampled and 200 ($\delta = 2\%$) posterior samples were retained. We illustrate our results for the facies model shown in Fig. 5. Fig. 10 compares estimates of the prior and posterior uncertainty. It can be clearly seen that modeling the direct relationship between $D$ and $H$ with DNNs and using it as the summary statistic reduces the prior uncertainty and captures the uncertainty around the reference case in both scenarios (compare to Fig. 6, summary statistics from unsupervised learning). The top right plot in Fig. 10 compares the true and DNN predicted values of the prior and posterior models. ABC selects those prior realizations as samples of the posterior for which $\hat{h}$ is nearest to $\hat{h}_{true} = S(\boldsymbol{d}_{obs})$ (black asterisk). Note that the total number of prior models evaluated by our approach, including the ones used for training purposes, amounts to 62000 in Scenario 1. As seen previously, features extracted using unsupervised methods are non-informative about NtG even with $2.02 \times 10^5$ prior model evaluations. As the final step in the workflow, posterior falsification, we evaluated our results on a test set of 2000 examples by the proposed method of posterior falsification (equation 12). As demonstrated in Fig. 11, the points along the diagonal indicate that equation 11 holds and the approximate posterior is not falsified. Our method makes reliable predictions of posterior uncertainty on the test set examples.

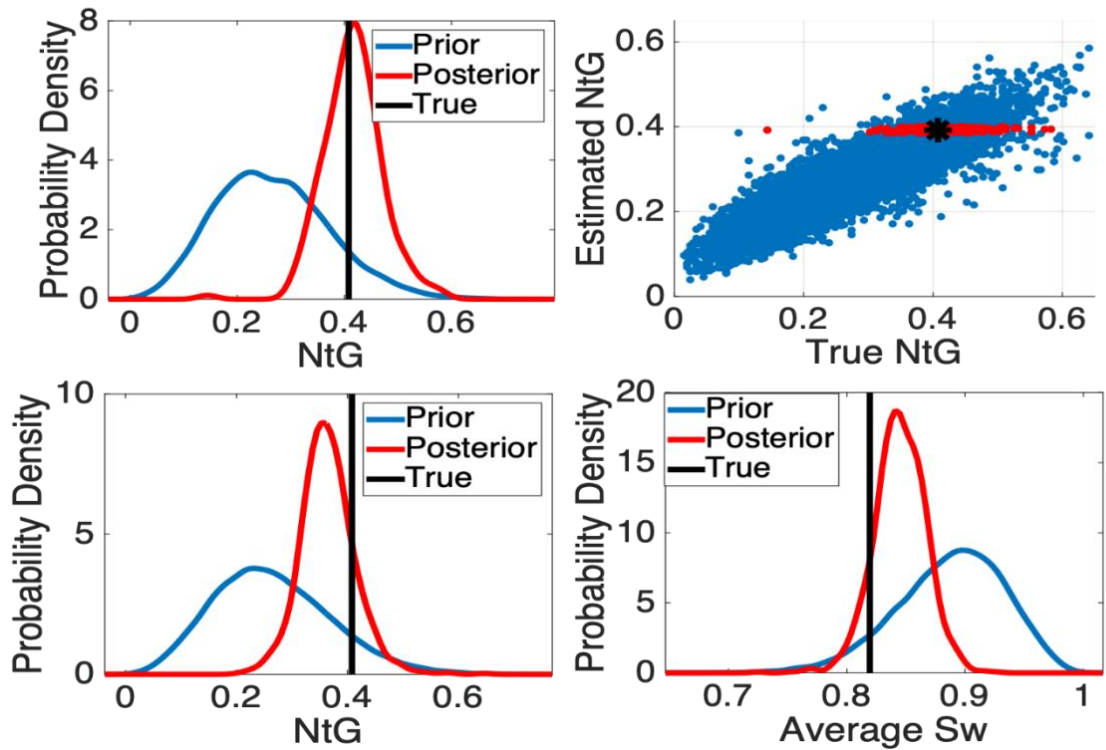

**Fig. 10** (Top: Scenario 1) Prior and posterior uncertainty of NtG estimated by seismic BEL using DNN as summary statistic on the left. The right plot compares the true and DNN estimated values of NtG for the prior and posterior models. (Bottom: Scenario 2) Prior and posterior uncertainties of NtG and average water saturations estimated by seismic BEL using DNN as summary statistic. The true facies model is shown in Fig. 5

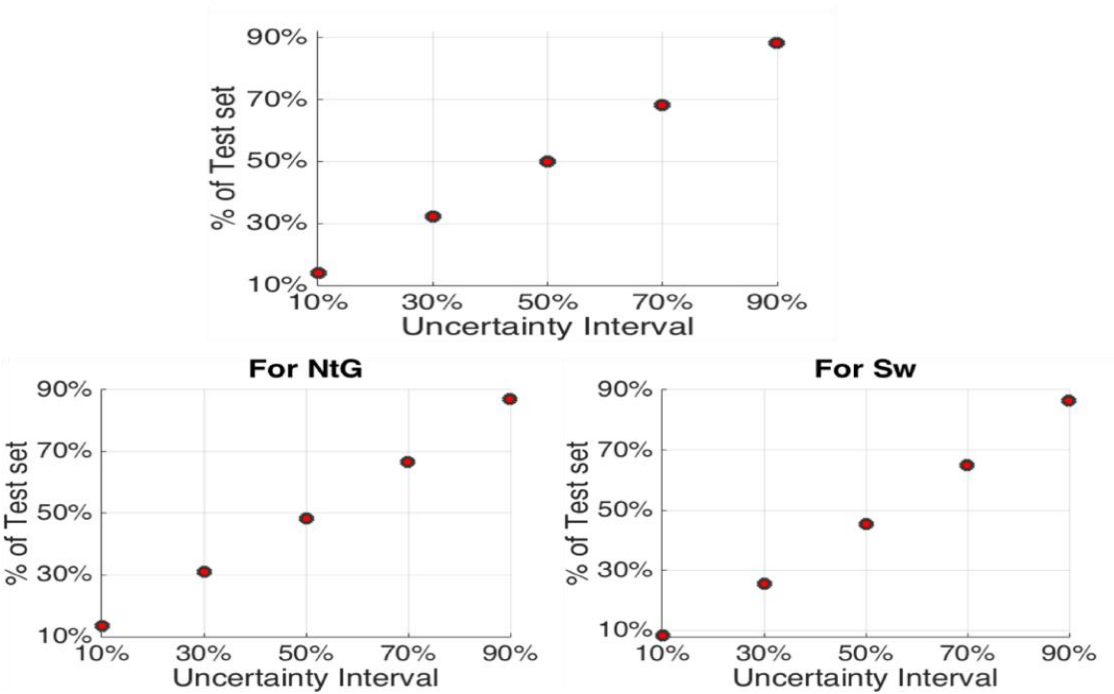

**Fig. 11** Percentage of test set examples for which corresponding posterior distributions captured the reference case in various uncertainty intervals (90% uncertainty interval denotes P5-P95). Results for scenario 1 and 2 are shown in the top and bottom rows respectively

# 5  Real case example from offshore Nile delta

We present a real case study from a producing field located in offshore Nile Delta. The target reservoir zone is located in the depth range of 2300 to 2700 meters and contains gas saturated sands as part of paleo slope-channel systems of Plio-Pleistocene age. The available dataset consists of 3D pre-stack and post stack seismic cubes and logs from several wells drilled in the area. Some studies on rock physics modeling and probabilistic seismic petrophysical inversion have been conducted in this field [1-3]. As identified in these studies, a critical limitation of the seismic data is that it has a dominant frequency of 15 Hz (Fig. 12) due to attenuation effects from several shallow gas clouds. Challenges exist with regards to the spatial characterization of the thin reservoir layers observed in the wells. Our goal in this study to exploit our seismic BEL framework to obtain uncertainty estimates of gross variability of the NtG of the sub-seismic reservoir layers in the field. We expect such estimates to prove useful in imposing informative constraints on future geo-modeling, reservoir characterization or forecasting endeavors.

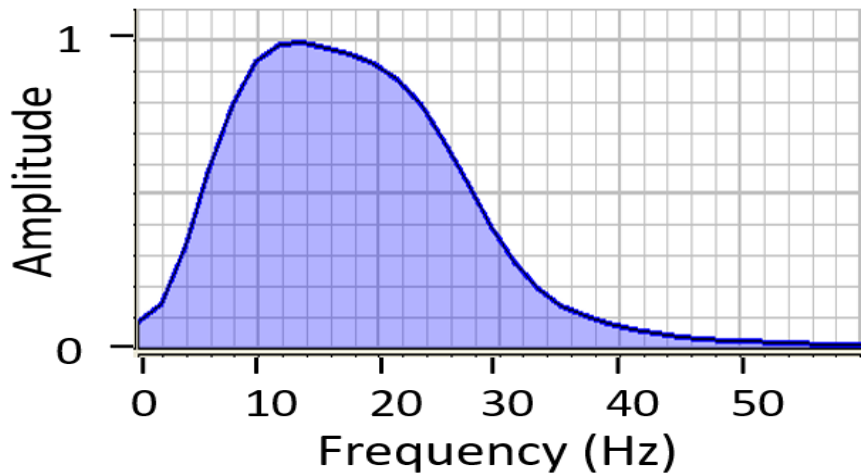

**Fig. 12** Amplitude spectrum extracted from the seismic data

We apply our seismic BEL workflow for a region with an approximate spatial extent of 3.6 km × 3.6 km. Our analysis will be focused on a window of approximately 200 m below the reservoir top horizon. Three wells are considered in the analysis. Well 1 is used in all components of the prior-building process. Well 2 lacks shear sonic information and is used only in derivation of the prior for facies model. Well 3 is kept blind. Fig. 13 illustrates elastic logs and interpreted facies log available in wells 1 and 3. The logs depict the presence of several sand layers, interbedded with shales. The sand layers in some cases are as thin as 1.2 meters. For the dominant seismic frequency of 15 Hz, these very thin sand layers are significantly below the seismic resolution ($\lambda/d \approx 136$; average velocity in the sonic logs $\approx$ 2450 m/s $\Rightarrow \lambda \approx$ 163 m). We will use our approach to estimate the depth averaged volume fraction of sand layers (NtG) across our zone of interest. Seismic data consists of near ($2^0$) and far ($30^0$) pre-stack angle gathers. Fig. 14 shows the root mean square (RMS) amplitudes extracted across the reservoir zone from the near and far angle gathers.

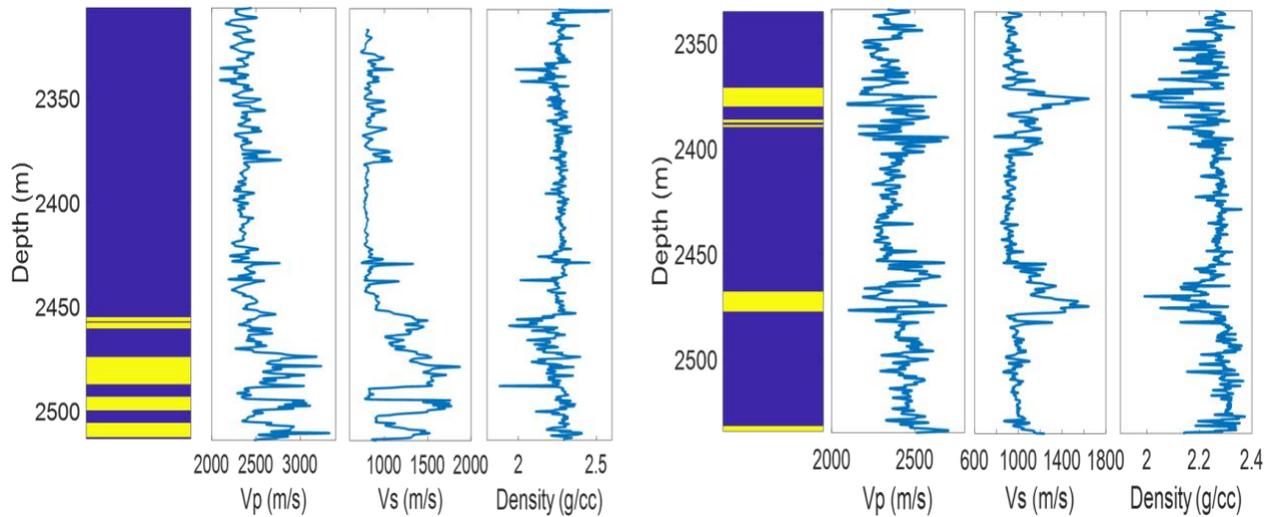

**Fig. 13** Interpreted facies (shale: blue, sand: yellow) and logged P-velocity, S-velocity and density profiles at well 1 (left) and 3 (right)

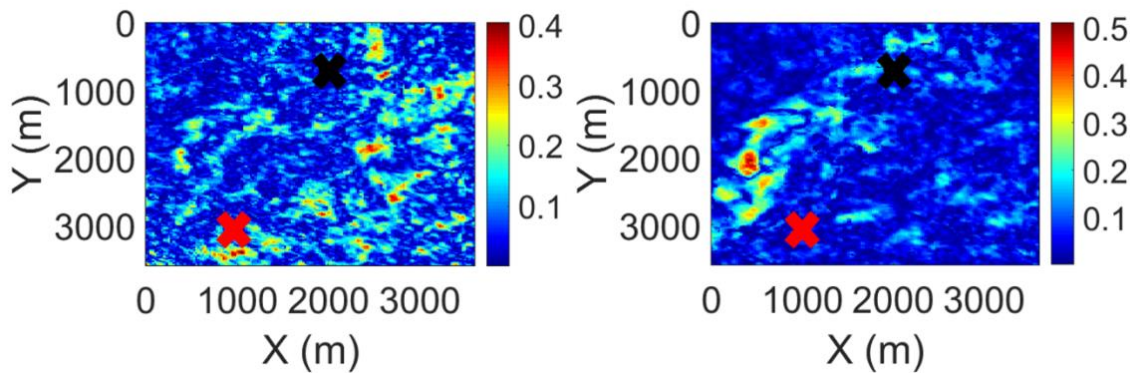

**Fig. 14** RMS near (left) and far (right) angle seismic amplitude maps extracted across the reservoir zone. Locations of well 1 (black cross) and well 3 (red cross) are shown as solid circles. Well 2 has the same well head as well 3

## 5.1 Prior specification and falsification

We use Markov chains and variograms (equivalently spatial autocorrelation functions) with a depth discretization of 15 cm to model the depth correlation of facies and elastic properties respectively. Lateral spatial correlation in the final estimates of NtG is imposed only by the trace-to-trace spatial correlation of the seismic data. Interpreted facies logs in wells 1 and 2 are used to derive the transition matrix for the Markov chain with 2 states representing sand and shale facies (Table 6). For these transition probabilities, the sand layers in Markov chain simulations will have $\lambda/d_{avg} \approx 30$ (average thickness of sand layers in well 1 is 6 meters and thus $\lambda/d_{avg} = 27$ at well 1). Fig. 15 shows some facies realizations obtained by unconditional simulations of the Markov chain with the specified transition matrix. The priors for elastic properties ($V_p$, $V_s$ and $\rho_b$) are modeled as multivariate Gaussian distributions. Compressional sonic, shear-sonic, and density logs at well 1 are used to model the spatial autocorrelation functions for each elastic property of each facies separately. Subsequently, realizations of $V_p$ are generated by Sequential Gaussian simulation (SGSIM [17]). Given an SGSIM realization of $V_p$, we generate corresponding realization of $V_s$ and $\rho_b$ by Sequential Gaussian co-simulation (COSGSIM). $V_p - V_s$ and $V_p - \rho_b$ cross-correlations of 80% and 20%, calculated with the logs at well 1, are imposed in co-simulations by the Markov1-type model [22]. Fig. 16 shows the facies-conditional prior distributions of $V_p$, $V_s$ and $\rho_b$. Corresponding realizations ($\in \mathbb{R}^{1349}$) of the full multivariate prior are shown in Fig. 15.

**Table 6** Transition matrix of the Markov chain in real case application

|       | Sand   | Shale  |
|-------|--------|--------|
| Sand  | 0.9725 | 0.0275 |
| Shale | 0.0068 | 0.9932 |

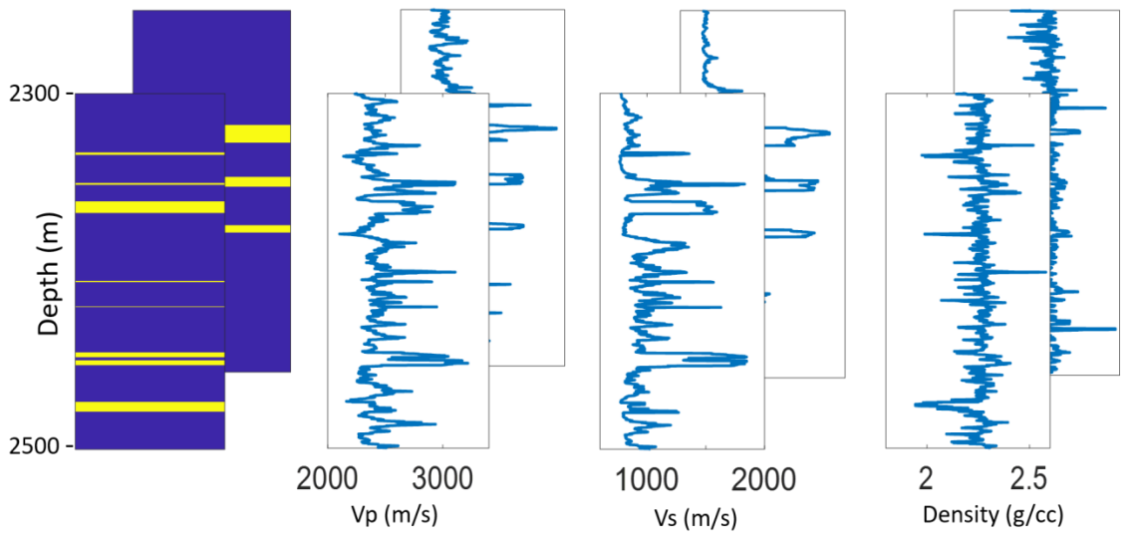

**Fig. 15** Realizations of facies (leftmost, shale: blue, sand: yellow) and elastic properties obtained by sampling the prior

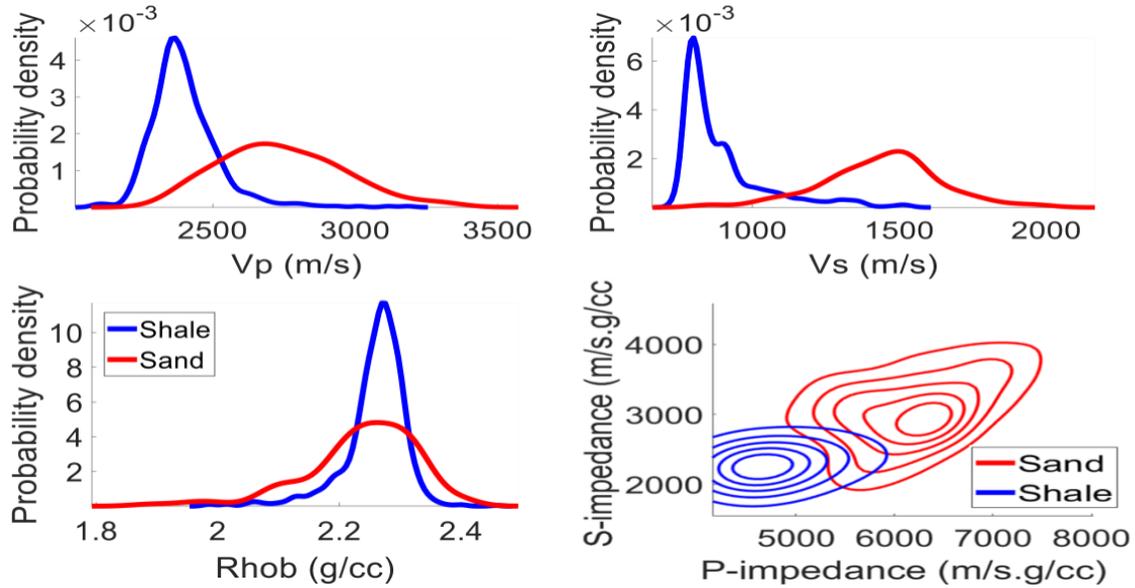

**Fig. 16** Kernel density estimates of prior facies-conditional distributions of $V_p$, $V_s$, $\rho_b$ and $I_p - I_s$ for real case application

Seismic angle gathers are generated using single-scattering forward modeling with the exact non-linear Zoeppritz equation, using wavelets extracted from the actual field data. Near and far angle wavelets, shown in Fig. 17, were extracted at well 1 by the coherency matching technique proposed by Walden and White [48]. Corresponding seismic-well ties for the near and far angle traces exhibit correlation coefficients of 86% and 94% respectively. The amplitudes of synthetic traces capture 70% and 68% of the variances of near and far angle traces respectively recorded at well 1. Corresponding amplitude errors can be attributed to the noise present in the data and physical effects not modeled in our forward modeling scheme. We assume that the probabilistic model for the data and forward-modeling noise ($\boldsymbol{E}_3$) is given by independent and identically distributed Gaussian distributions having zero mean and 30% and 32% (SNR ≈ 3) variances for near and far angles respectively. It is desirable for the DNN to learn this underlying noise distribution and account for it while making predictions. We propose to achieve this objective by generating samples from the noise distribution and add it to the training examples of the data according to equation 3.

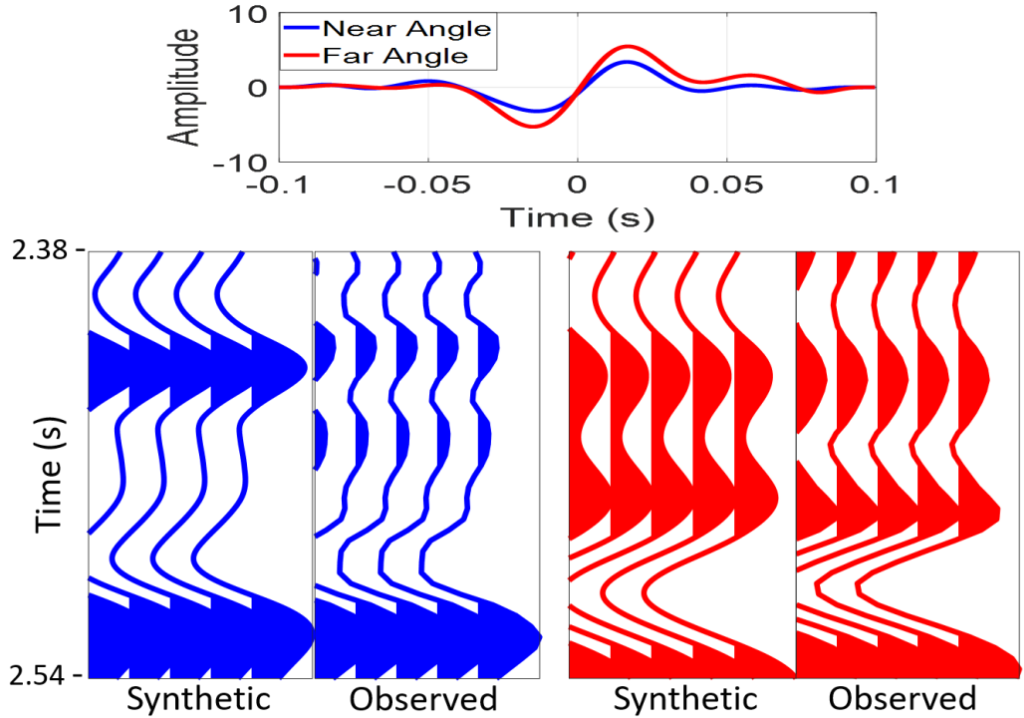

**Fig. 17** (Top) Wavelets extracted for seismic forward modeling. (Bottom) Seismic-well tie at well 1. The near angle traces are shown in blue and far angle traces in red

For real case applications, it is essential to validate the consistency of the defined prior distributions with observed data. We ensure that the modeled prior distributions for elastic properties are consistent with well logs by comparing autocorrelation functions of 5000 prior realizations against that of the well logs as shown in Fig. 18.

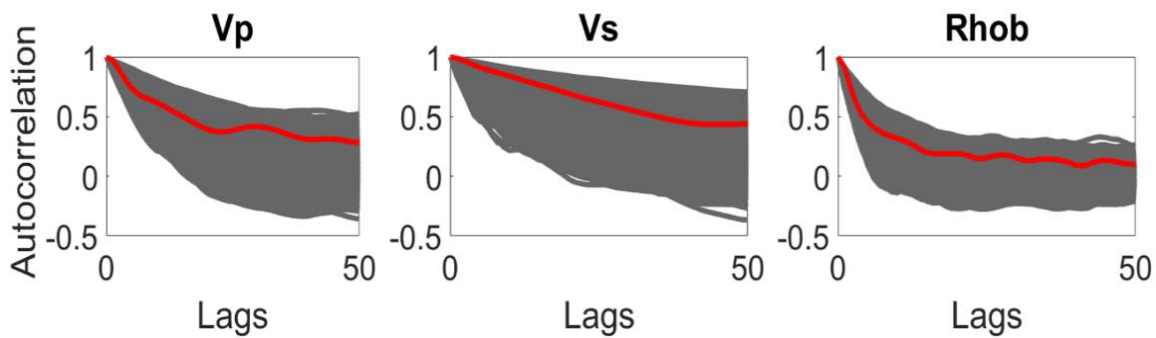

**Fig. 18** Autocorrelation functions of the prior realizations (gray) and well-log (red) for elastic properties

To compare the observed and forward modeled seismic traces, we employ the prior falsification approach described in the theory section. Inconsistent priors will be falsified by outlier detection using Mahalanobis distances [35]. To identify outliers in a given set of $n$-dimensional samples $\{x_i \in \mathbb{R}^n; i = 1,..,l\}$, we compute the Mahalanobis distance

$$d^M = \sqrt{(x - \hat{\mu})^T \hat{\Sigma}^{-1}(x - \hat{\mu})}, \quad (16)$$

where $\hat{\mu}$ and $\hat{\Sigma}$ are an estimate of the data mean and covariance. Note that setting $d^M$ to a constant value corresponds to defining an ellipsoid in the multivariate space centered at $\hat{\mu}$. The approach for outlier detection is to define a threshold $\epsilon$ such that $x_i$s located outside the ellipsoid defined by $\epsilon$ are deemed as outliers. The threshold is generally assigned as

$$\epsilon = \sqrt{\chi^2_{n,0.975}} \quad (17)$$

[35], where $\chi^2_{n,0.975}$ is the 97.5% quantile of the chi-square distribution with $n$ degrees of freedom. Note that if the underlying data distribution is a multivariate Gaussian, then the probability that $d^{M^2} > \chi^2_{n,0.975}$ is $(1 - 0.975)$. Thus, in a large dataset, there is a finite probability that a few samples from the prior distribution may have a Mahalanobis distance exceeding the threshold.

We present the prior falsification analysis with a set of 5000 prior samples of $D$ ($\in \mathbb{R}^{760}$) and seismic traces acquired at 64 locations in the area of interest. The 64 locations were selected by coarsely sampling the original seismic survey grid every 500 meters along inline and crossline directions. Note that we have approximately 42000 seismic traces in the area of interest. We compute Mahalanobis distances in a reduced dimensional space obtained using PCA and multi-dimensional scaling (MDS [12]). 5064 seismic traces (5000 prior samples plus 64 field seismic traces) were used to identify 285 principal components which retain 90% of the original variance. Subsequently, we use MDS which facilitates dimension reduction while preserving some measure of dissimilarity in the original data and MDS space. We performed MDS on the 5064 traces represented in the 285-dimensional principal component space, using Euclidean distance between the principal component scores as a measure of dissimilarity. In Fig. 19, we show the prior and acquired seismic traces in different bivariate plots of the first three MDS dimensions. It can be ascertained by visual inspection that none of the field seismic traces feature as outliers in the first three MDS dimensions. Note however that the mutual distances between the traces in the trivariate MDS space has a correlation coefficient of approximately 65% with corresponding distances in the 285-dimensional principal component space.

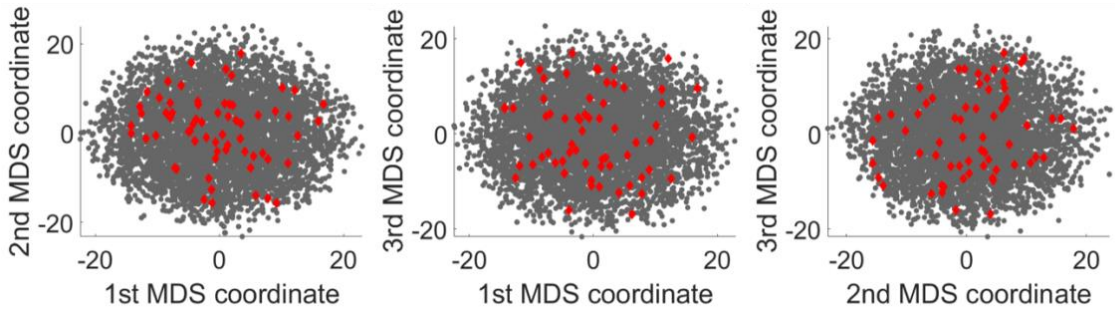

**Fig. 19** 5000 prior samples of forward modeled seismic data (gray circles) and seismic traces extracted at 64 locations from the acquired pre-stack seismic cube (red diamonds) plotted along various MDS dimensions

Going beyond a qualitative visual check, we perform Mahalanobis distance based outlier detection in the first 12 MDS dimensions. We select 12 MDS dimensions as the 12 dimensions exhibit 95% correlation coefficient between distances in MDS and principal component space. We estimate $\widehat{\boldsymbol{\mu}}$ and $\widehat{\boldsymbol{\Sigma}}$ using two approaches: 1) The classical approach where $\widehat{\boldsymbol{\mu}}$ and $\widehat{\boldsymbol{\Sigma}}$ are assigned as the sample mean and covariance, and 2) robust estimates of $\widehat{\boldsymbol{\mu}}$ and $\widehat{\boldsymbol{\Sigma}}$ obtained by the minimum covariance determinant estimator [36]. This approach addresses the masking effect caused by presence of outlier clusters in the samples which may bias the sample mean and covariance in the direction of the outliers. In Fig. 20, we plot the classical and robust Mahalanobis distances of the 5064 seismic traces estimated in the first 12 MDS dimensions. The Mahalanobis distances for a few prior samples fall above the threshold value $\sqrt{\chi^2_{12,0.975}}$ using the classical (6 samples above the threshold) and robust (37 samples above the threshold) estimates. More importantly, it can be seen that all the selected field seismic traces are within the defined threshold value, thus indicating that the stated priors are consistent with the field seismic traces. The prior is not falsified.

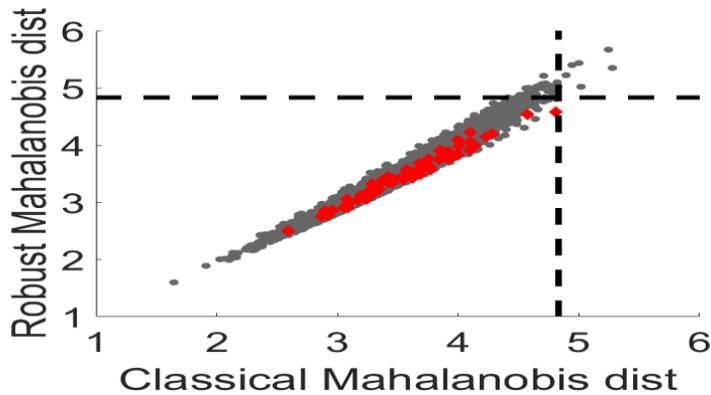

**Fig. 20** Classical Mahalanobis distances of the prior (gray circles) and field (red diamonds) seismic traces plotted against the robust Mahalanobis distances computed in 12 MDS dimensions. Shown in dotted black lines are the values for $\sqrt{\chi^2_{12,0.975}}$

## 5.2  Informative statistics, Approximate Bayesian computation and posterior falsification

Subsequent to the prior falsification, the next steps involve training the DNN that will be used to estimate the informative statistic, followed by ABC and posterior falsification. A training set for the DNN training is generated by sampling from the prior distribution. The DNN architecture is similar to the one for the synthetic example. Specific details about the optimized DNN hyper-parameters and training are presented in Table 3 and Table 4 respectively. Fig. 21 shows the performance of the network on training and validation sets. Even though the DNN has achieved good generalization ability, the reliability of its predictions has deteriorated in comparison with the results of the synthetic example presented previously. Possible reasons which render this example more challenging include: 1) Seismic wavelength to layer thickness ratio is much larger 2) Significantly more noise is present in the real case.

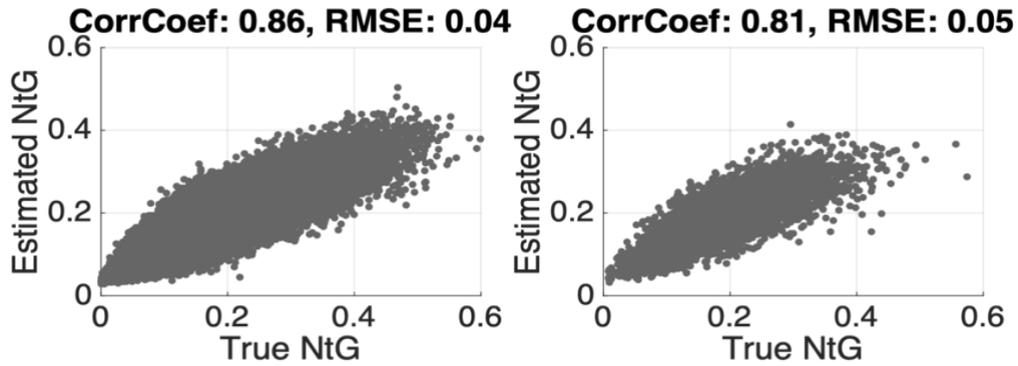

**Fig. 21** Plot of actual NtG against the NtG estimated by the trained network for the training (left) and validation (right) set for the real case application

Once we have the informative statistics from the trained DNN, the next steps are inference using ABC and posterior falsification. Posterior falsification analysis is performed with a separate test set of 5000 examples. ABC rejection sampling is performed with a set of 30000 prior models and the approximate posterior for each test set example is estimated with 300 samples (threshold $\delta = 1\%$ of 30000). Fig. 22 indicates that the posterior estimated with the DNN as summary statistic and $\delta = 1\%$ is a reasonable approximation to the desired posterior distribution.

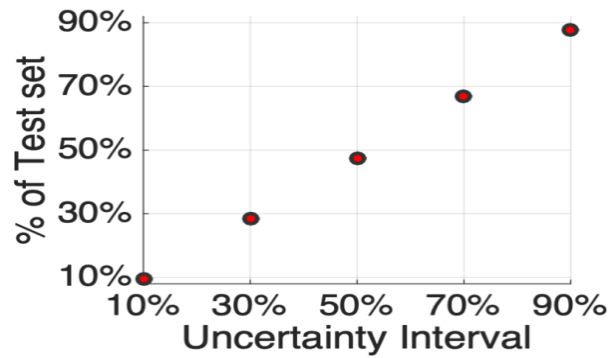

**Fig. 22** Results of posterior falsification for the real case application

Now we can apply the DNN and ABC inference to all the traces in the seismic cube. The seismic gather at each spatial location in the area of interest is input to obtain the NtG prediction from the DNN as shown in Fig. 23. Posterior distributions were estimated at each location using our approach and corresponding 90% uncertainty intervals are also shown. It can be observed that posterior uncertainty intervals are somewhat correlated with the NtG predictions from the DNN. Regions of higher predicted NtG also have greater posterior uncertainty. This result is explicated in Fig. 24 where we compare the prior and posterior distributions for three test set examples with increasing NtG values. The trained network generally predicts examples with higher NtG values with lower confidence. Consequently, the accepted posterior models have larger scatter of true NtG values around the DNN estimated values. To validate our results, we present the estimates of posterior uncertainty at well 1 and blind well 3 in Fig. 25. It can be observed that uncertainty predictions by our approach has successfully captured the true NtG interpreted at these wells.

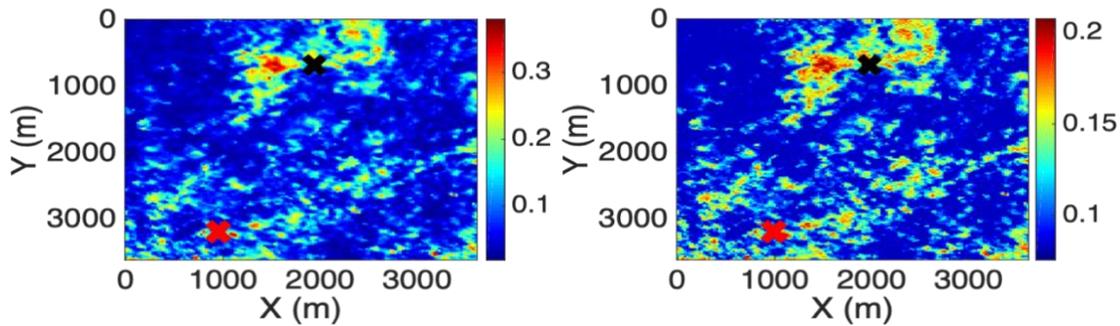

**Fig. 23** (Left) NtG predicted by the DNN using near and far angle seismic traces. (Right) 90% uncertainty interval of the estimated posterior distributions. Locations of well 1 (black cross) and well 3 (red cross) are shown as solid circles

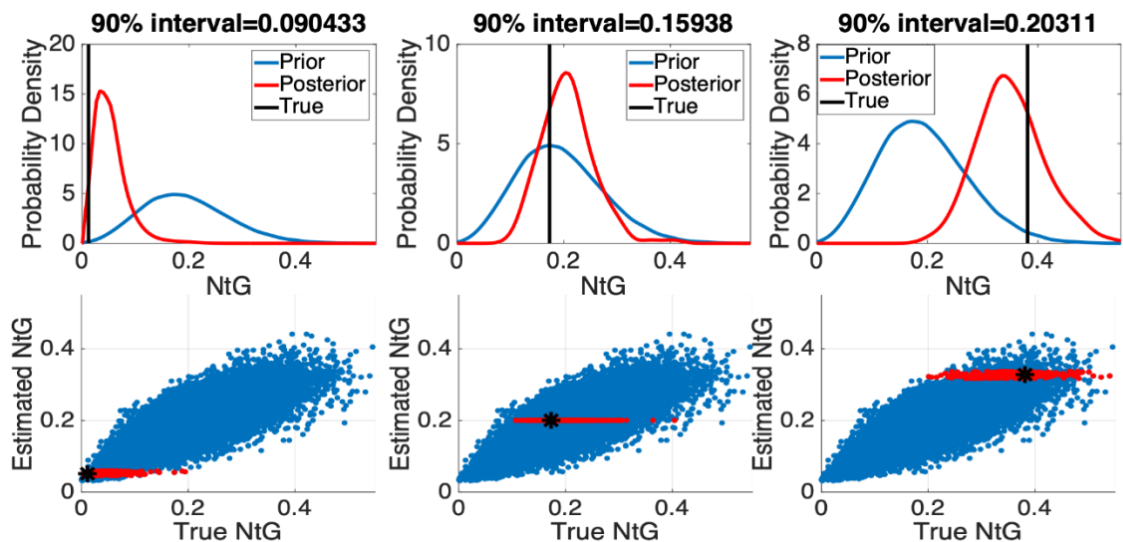

**Fig. 24** (Top) Prior and posterior distributions for three test set examples with increasing NtG values from left to right. The 90% posterior uncertainty interval (P5-P95) is stated on the top.

(Bottom) Corresponding scatter plots of true and DNN estimated NtG values. Estimated posterior uncertainty intervals change in proportion to the confidence we have on the predictions by the trained network

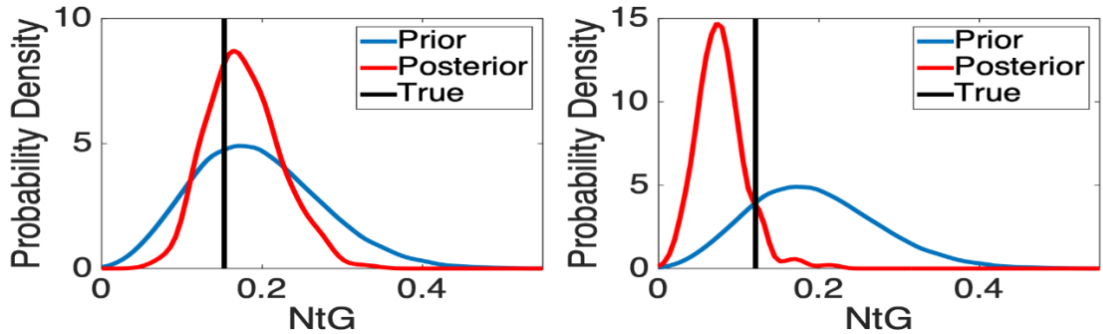

**Fig. 25** Approximate posterior distributions of NtG estimated at well 1 (left) and blind well 3 (right). The true NtG values estimated using the facies logs are shown in black

## 6  Discussion and Conclusions

Though seismic data are the elastic response of the subsurface, estimation of low-dimensional non-elastic quantities from seismic data does not necessitate solving a demanding seismic inverse problem. In this paper, we presented an evidential learning framework for performing seismic estimation of any low-dimensional earth property implicitly related to seismic data. Our approach employs Approximate Bayesian Computation for performing inference and uncertainty quantification with any desired statistical learning model trained to encapsulate this implicit relationship.

Employing statistical models, in conjunction with physical models, for solving estimation problems comes with several pitfalls. Inconsistency of stated prior distributions with acquired data is a major limitation in our approach because these priors are also employed to generate training sets for the statistical learning method. Any formulated prior distributions should be subjected to appropriate prior falsification tests. Another critical aspect is that the trained statistical model should have good generalization ability to examples not presented during training. A model which is overfitting the training dataset will significantly bias the predicted estimates of uncertainty. We discussed employing validation sets during training and performing posterior falsification subsequent to training as precautionary measures against this pitfall.

Notwithstanding the above limitations, a crucial advantage of our approach is the way we approach sampling of probability distributions on uncertain parameters. Generation of the training set is accomplished by random sampling of prior distributions. ABC inference of low-dimensional target variables can also be performed efficiently with random sampling, subject to the condition that summary statistics are sufficiently informative. This offers the flexibility to incorporate prior distributions of desired complexity into the analysis. Random sampling is generally an easier problem than designing strategies for smart exploration of complex high-dimensional conditional probability spaces, often required for solving traditional Bayesian inverse problems.

Major factors controlling the computational costs of the approach are: 1) Number of examples required to train the statistical model without overfitting 2) Number of prior examples required to perform ABC. 3) The degree to which the model is informative on target variables. The first factor will generally depend on the nature of the relationship between the data and target variables, dimensionality of the problem space and the employed statistical model. The second factor is predominantly linked to the dimensionality of the data or target variables since high dimensional variables will render the ABC rejection sampling procedure highly inefficient. An informative summary statistic can be very effective in limiting these costs and thus special attention should be directed towards selecting an appropriate statistic. Unsupervised learning models are generally easier to be trained and less prone to overfitting than their supervised counterparts. Supervised models, however, facilitate directly encoding the relationship of data and target variables into the summary statistic.

We demonstrated the efficacy of our approach on a synthetic example and real case study of seismic estimation in reservoir intervals with sub-seismic sand layers. We explored two different approaches to estimate summary statistics. In the unsupervised approach we analyzed the applicability of PCA, wavelet thresholding and autoencoders. For supervised learning we used deep neural networks to extract data summary statistics for estimation of NtG and average fluid saturation. Compared to the unsupervised methods, the supervised deep neural network architecture was found to be more effective in extracting highly informative summary statistics on the target properties. We successfully applied our method to estimate the NtG and average fluid saturation uncertainty of sub-resolution thin-sand using near and far-offset seismic waveforms, without requiring an explicit inversion.

## Acknowledgements

This work is supported by the sponsors of the Stanford Center for Earth Resources Forecasting (SCERF), and support from the Dean of the Stanford School of Earth, Energy, and Environmental Sciences, Professor Steve Graham. We thank Edison E&P for providing the dataset for the real case application. We offer special thanks to Dr. Fabio Ciabarri for his advice and assistance regarding various aspects of the real dataset. We would also like to thank Professor Jef Caers for insightful discussions regarding the methodology.